# A Connection between Spectral Width and Energetics As Well As Peak Luminosity in *Fermi* Gamma-Ray Bursts

Z. Y. Peng[1], X. H. Zhao[2], Y. Yin[3], and D. Z. Wang[1]
[1] College of Physics and Electronics, Yunnan Normal University, Kunming 650500, People's Republic of China; pengzhaoyang412@163.com
[2] Yunnan Observatories, Chinese Academy of Sciences, 396 Yangfangwang, Guandu District, Kunming, 650216, People's Republic of China
[3] Department of Electrical Engineering, Liupanshui Normal University, LiuPanShui 553004, People's Republic of China

## Abstract

We have revisited the spectral width in the $EF_E$ spectrum of gamma-ray bursts with the BEST peak flux P and time-integrated F spectral data provided by the *Fermi* GBM Burst Catalog. We first compute the BEST spectral widths to compare with some typical physics models. Our analysis results consist with the previous results: blackbody emission alone cannot explain the observed spectrum and most of the observed spectra cannot be interpreted by the synchrotron radiation. We then check the correlations between the spectral width and the observable model-independent burst properties of duration, fluence, and peak flux and find that positive correlations exist between them for both the P and F spectra. Moreover, the short burst appears to extend the correlation found for the long burst. We further demonstrate that these positive correlations also exist in the cosmological rest frame; that is, the spectral width correlates with the isotropic-equivalent energy $E_{\rm iso}$ as well as the isotropic-equivalent peak luminosity $L_{\rm iso}$ for different energy bands and timescales. Our results show that the wider bursts have larger energy and luminosity. Moreover, short bursts would appear to extend this trend qualitatively. Taking the Amati relation into account, we tend to believe that the spectral shape is related to energy and luminosity.

## 1. Introduction

The emission mechanisms in gamma-ray burst (GRB) prompt emission are still poorly understood although nearly 50 years have passed. The analysis of the GRB prompt emission, especially for the spectroscopy, provides us with valuable clues about the underlying processes giving rise to the phenomenon. The temporal profiles of GRBs are very diverse in morphology but the overall spectral shape is similar, which can be fitted with a single simple Band model (Band et al. 1993) or the cutoff power-law model. Because the Band model is an empirical function, the spectral parameters (the power-law indices and the peak energy in the $\nu f_\nu$ spectrum) are then used to infer the GRB emission and particle acceleration mechanisms. For example, the lower-energy index is usually compared to the slopes of various radiation models, leading to the discovery of the so-called line-of-death problem (e.g., Katz 1994; Preece et al. 1998; 2002) for the synchrotron theory.

It is generally believed that GRBs are divided into short- and long-duration classes based on $T_{90}$ where all the bursts are likely to be separated at about 2 s (Kouveliotou et al. 1993). Many researchers come to the conclusion that different classes might have different progenitors: short bursts are produced in the event of binary neutron star or neutron star−black hole mergers, whereas long bursts are thought to be the massive star collapses (e.g., Paczynski 1986, 1998; Eichler et al. 1989; Woosley 1993; MacFadyen & Woosley 1999). Many studies focus on the classification of GRBs (e.g., Horváth 1998, 2002; Lü et al. 2010; Qin & Chen 2013; Tarnopolski 2015; Horváth & Tóth 2016) but no consensus on this issue has been made.

Recently, Yu et al. (2016) focused on the explanation of the spectral peaks or breaks of the GRB prompt emission phase by studying the sharpness of the 1113 time-resolved prompt emission spectra of GRBs. They obtained a measure of the curvature of time-resolved spectra and then compared the curvature directly to theory. It is found that optically thin synchrotron radiation cannot fully explain the spectral peaks or breaks. While Axelsson & Borgonovo (2015, hereafter Paper I) concentrate on the spectral width only depending on Band spectral parameters $\alpha$ and $\beta$ using the full-width-at-half-maximum measurement of GRB prompt emission spectra. The peak flux data from the BATSE 5B GRB spectral catalog (Goldstein et al. 2013) and 4 yr *Fermi* GBM spectral catalog (Gruber et al. 2014, hereafter Paper II) are adopted to compute the spectral widths. They found that the distribution of the spectral widths comes from the two instruments that are fully consistent. However, the median of the widths from long and short bursts are significantly different. They have shown that a significant fraction of bursts (78% for long and 85% for short GRBs) could not be explained by a Maxwellian population-based slow-cooling synchrotron function via comparing the known emission mechanisms.

Several articles by Guiriec et al. (2010, 2011, 2013, 2015a, 2015b, 2016a, 2016b) have shown that the prompt emission is composed of a superposition of several components; one of them is a thermal spectral shape matching with photospheric emission. They showed that the shape of the nonthermal component can be dramatically different from the Band-only fits in the context of their multicomponent model, especially the value of $\alpha$, and, therefore, the curvature of the nonthermal spectrum is usually broader too. They find that in the context of their multicomponent model, the values of $\alpha$ do not change much during the burst emission episode and that these values seem to be the same for all short and long GRBs. Moreover, these values seem to be compatible with either slow or fast cooling synchrotron emission.

However, Paper I only employed the peak flux data fitted by the Band model. In fact, as mentioned by the authors, most GRBs are adequately fitted to COMP and the extra parameter $\beta$ of the Band function is not required to get a good fit (e.g., Preece et al. 2016). And only a small fraction of bursts are best fitted by the Band model from Paper II. Therefore, we wonder if the computed spectral widths only from Band are consistent with those of the BEST model from Paper II. In addition, Paper I pointed out that the median widths of spectra from long and short GRBs are significantly different and the width does not correlate with duration. Since the widths from long and short bursts are different we suspect that the spectral width may be correlated with duration and other physics quantities. These ideas motivate our investigations in this paper. We first compute the BEST spectral widths to compare with some typical physics models. Then we investigate the intrinsic connection between the GRB spectral width and GRB total radiated energy as well as peak luminosity. In Section 2, we present a sample description and data analysis. The results of the analysis are given in Section 3. In Section 4, we investigate the dispersion analysis and the effect of the GBM instrument on the spectral width. Discussion and conclusions are presented in the last section.

## 2. Sample Selection and Data Analysis

Paper I employed two instrument data to study the spectral width and found that there are not significant differences between the medians of the GBM sample against the larger BATSE sample; it also confirmed that the width distribution is not dependent on a given instrument. Therefore, in this paper, we only consider the GBM sample. An enormous spectral catalog detected by the *Fermi*/GBM in the energy range of 8 keV–40 MeV is provided by the *Fermi* GBM burst catalog published at HEASARC (which is available at https://heasarc.gsfc.nasa.gov/W3Browse/fermi/fermigbrst.html). It significantly expands our understanding of the physical properties and characteristics of GRBs. The *Fermi* GBM burst catalog provides two types of spectra, the time-integrated spectral fits (F spectra) and spectral fits at the brightest time bin (P spectra). They employ four different spectral models to fit the data, respectively. These models include a single power law (PL), the Band GRB function (BAND), an exponential cutoff power law (COMP), and a smoothly broken power law (SBPL). We use all GRB spectral data between 2008 July and 2018 October from the catalog.

Paper I presented the width distributions of the long and short GRBs, respectively. But the spectral widths are only computed with the Band model. However, the actual spectrum might not be best described by a Band function. Moreover, many studies have pointed out that one should not only assume a Band spectrum when performing spectral analysis of GRBs and should always try different fit functions and compare the fit statistics to find the best description to the data (e.g., Paper II; Giblin et al. 1999; Guiriec et al. 2010, 2011, 2013, 2015a, 2015b, 2016a, 2016b; González et al. 2012; Sacahui et al. 2013). Therefore, we compute the spectral widths with the BEST model parameters in Paper II. To minimize the selection bias, we select all of the bursts from the BEST models with curvature shapes, which includes the BAND, COMP, and SBPL models. There are two kinds of spectral widths. The first width is the relative width defined by Paper I, which is rewritten as follows:

$$W = \log\left(\frac{E_2}{E_1}\right), \quad (1)$$

Table 1
A List of the Sample Size of the Three Models

| | P Spectra | | | F Spectra | | |
|---|---|---|---|---|---|---|
| | Long GRBs | Short GRBs | Entire Bursts | Long GRBs | Short GRBs | Entire Bursts |
| BAND | 59 | 1 | 60 | 166 | 2 | 168 |
| SBPL | 29 | 1 | 30 | 86 | 4 | 90 |
| COMP | 914 | 136 | 1050 | 1168 | 205 | 1373 |
| All | 1002 | 138 | 1140 | 1420 | 211 | 1631 |

where $E_1$ and $E_2$ are the lower and upper energy bounds of the full width at half maximum of the $EF_E$ versus E spectra, respectively.

The second width is the absolute width, which can be written as follows:

$$W_{ab} = \log(E_2 - E_1). \quad (2)$$

In this paper we mainly adopt the relative width defined by Paper I. Different from Paper I, we only require $\alpha > -2.0$ and $\beta < -2.05$ to minimize selection effects due to parameter limit. The uncertainty for each burst width is estimated by using Monte Carlo methods.

We divide the entire burst set for two types of spectra into two subsets, the long and short burst sets. The bursts with $T_{90} > 2$ s are classified as long bursts; otherwise, the bursts are short bursts. Note that for the P spectra the time selection performed by Paper II is 1.024 and 0.064 s for the long and short bursts, respectively. We analyze the entire burst set, as well as long and short burst sets, with the BEST model separately for the P spectra and F spectra. The number of various models including two types of bursts for the P and F spectra is listed in Table 1. There are 1140 P spectra and 1631 F spectra, in which there are 138 and 211 short bursts for the P and F spectral data, respectively. The number of the P spectra is evidently less than that of the F spectra due to less photon fluence accumulation and more GRBs of the P spectra, and the spectra are best fitted by the PL model.

## 3. Analysis Results

We first check the width distributions of two classes of spectra. Then we wonder if some correlations exist between the spectral widths and some physics quantities.

### 3.1. The Distribution of the Burst Spectral Width

The spectral widths fitted by different models are very different. As shown by Table 2, the median widths of the SBPL and BAND model are much larger than that of the COMP model, especially for the F spectra. The width distributions for the long and short bursts are demonstrated in Figure 1 and the characteristics are summarized in Table 3. We find from Figure 1 and Table 3 for the long burst of two types of spectra that (1) the widths range from $0.62 \pm 0.04$ to $4.92 \pm 0.66$ for the P spectra and from $0.67 \pm 0.04$ to $4.91 \pm 1.26$ for the F spectra; (2) the distribution peaks at <1 and >1 for the P spectra and the F spectra—for both of the spectra, there is a very small fraction of bursts extending toward larger widths; and (3) the corresponding median values of $W$ are $0.96 \pm 0.04$ and $1.09 \pm 0.08$ for the P spectra and the F spectra, respectively.

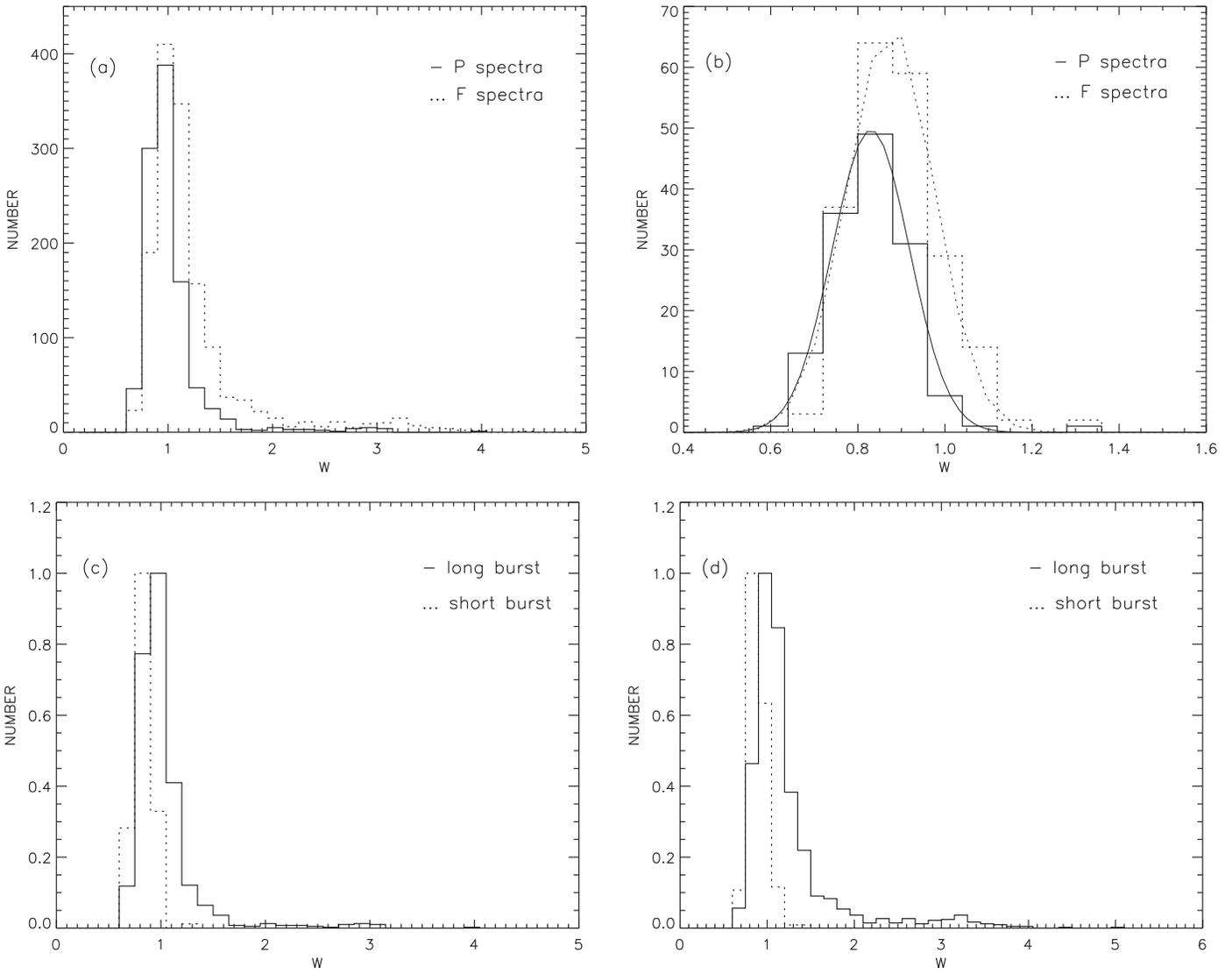

**Figure 1.** Distributions of the spectral width $W$ for (a) the long and (b) the short burst set and the comparison of the spectral width distribution between the long and short burst for (c) the P spectra and (d) the F spectra.

**Table 2**
Characteristics of the Spectral Width Distributions of the Two Types of Spectra, Separated for the BAND, COMP, and SBPL Model

| | P Spectra | | | F Spectra | | |
|---|---|---|---|---|---|---|
| | BAND | COMP | SBPL | BAND | COMP | SBPL |
| Sample Size | 60 | 1050 | 30 | 168 | 1373 | 90 |
| Median | 1.27 ± 0.17 | 0.93 ± 0.11 | 1.49 ± 0.17 | 1.72 ± 0.07 | 1.02 ± 0.02 | 1.89 ± 0.27 |
| Minimum | 0.76 ± 0.04 | 0.62 ± 0.03 | 0.82 ± 0.32 | 0.87 ± 0.02 | 0.66 ± 0.04 | 0.82 ± 0.02 |
| Maximum | 4.04 ± 0.58 | 2.25 ± 0.20 | 3.08 ± 0.39 | 4.98 ± 0.90 | 2.66 ± 0.46 | 3.96 ± 0.60 |

**Table 3**
Characteristics of the Spectral Width Distributions of the Two Types of Spectra, Separated for the Long and the Short GRBs

| | P Spectra | | | F Spectra | | |
|---|---|---|---|---|---|---|
| | Long GRBs | Short GRBs | Entire Bursts | Long GRBs | Short GRBs | Entire Bursts |
| Number | 1002 | 138 | 1140 | 1420 | 211 | 1631 |
| Median | 0.96 ± 0.09 | 0.82 ± 0.10 | 0.94 ± 0.08 | 1.09 ± 0.05 | 0.88 ± 0.09 | 1.05 ± 0.04 |
| Minimum | 0.62 ± 0.03 | 0.64 ± 0.04 | 0.62 ± 0.03 | 0.66 ± 0.04 | 0.70 ± 0.02 | 0.66 ± 0.04 |
| Maximum | 4.04 ± 0.58 | 1.33 ± 0.37 | 4.04 ± 0.58 | 4.98 ± 0.90 | 3.72 ± 0.14 | 4.98 ± 0.90 |

**Table 4**
Spearman Rank Correlation Analysis Results of the Spectral Widths and the Other Parameters for All the Models

| | P Spectra | | | F Spectra | | |
|---|---|---|---|---|---|---|
| Parameter | Entire GRBs | LGRBs | SGRBs | Entire GRBs | LGRBs | SGRBs |
| $W - T_{90}$ | 0.29 ($6.46 \times 10^{-23}$) | 0.13 ($6.85 \times 10^{-5}$) | 0.07 (0.42) | 0.40 (0.00) | 0.23 ($8.5 \times 10^{-19}$) | 0.22 ($1.42 \times 10^{-3}$) |
| $W - $ flue | 0.29 ($2.20 \times 10^{-23}$) | 0.17 ($3.71 \times 10^{-8}$) | −0.23 (0.007) | 0.47 (0.00) | 0.37 (0.00) | 0.14 (0.04) |
| $W - f_{1024}$ | 0.33 ($2.05 \times 10^{-31}$) | 0.29 ($1.24 \times 10^{-21}$) | −0.06 (0.47) | 0.45 (0.00) | 0.41 (0.00) | 0.36 ($6.17 \times 10^{-8}$) |
| $W - f_{256}$ | 0.24 ($1.31 \times 10^{-16}$) | 0.29 ($1.99 \times 10^{-21}$) | −0.09 (0.31) | 0.33 ($2.78 \times 10^{-43}$) | 0.40 (0.00) | 0.31 ($6.34 \times 10^{-6}$) |
| $W - f_{64}$ | 0.19 ($3.19 \times 10^{-11}$) | 0.29 ($1.39 \times 10^{-20}$) | −0.01 (0.23) | 0.28 ($3.30 \times 10^{-31}$) | 0.38 (0.00) | 0.23 ($8.37 \times 10^{-4}$) |

While for the short burst of two types of spectra (1) the widths range from 0.64 ± 0.04 to 1.60 ± 0.05 for the P spectra and from 0.70 ± 0.02 to 1.44 ± 1.16 for the F spectra; (2) the distribution peaks <1 for both spectra and the corresponding mean values and standard deviations are around 0.83 and 0.10 for the P spectra and 0.87 and 0.09 for the F spectra, respectively; (3) the corresponding median values of $W$ are 0.82 ± 0.06 and 0.88 ± 0.03 for the P spectra and the F spectra, respectively. It is found that the widths of the F spectra shift to a larger values relative to the P spectra for both the long bursts and the short bursts.

When comparing our results of the P spectra with those of Paper I we find that, for both the long and short bursts, the medians are much smaller than those of Paper I. That is, the spectral width tends to be a smaller value when we adopt the BEST model, which also reveals that the true spectral width is smaller than that of Paper I because the BEST model can better represent the GRB spectra. While for the F spectra, both of the median values of the long and short bursts are very close to those of the P spectra provided by Paper I (see Table 3).

When comparing the long and short bursts for two types of spectra from Figure 1 and Table 3 it is found that the peak value and the median of the short burst widths are evidently smaller than those of the long burst for both spectra. This is consistent with the result of Paper I. Whether or not only the median value is different for the two types of bursts is the question. We do a K-S test to check this, and it is found that the significance probabilities and $D$ values are $5.14 \times 10^{-24}$, 0.47 and $5.69 \times 10^{-43}$, 0.54 for the P and the F spectra, respectively. Therefore we can conclude that there is a highly significant difference between the two types of GRBs. Moreover, the difference of the F spectra is more significant. However, there is substantial overlap between the long burst and short burst (see, Figure 1 and Table 3). Therefore the distributions of the two types of bursts are perfectly compatible when taking into account the variances of the distributions.

We also compare the width distributions between the F spectra and the P spectra. For the short burst, the probability ($5.32 \times 10^{-5}$) and $D$ value (0.25) reveal that the distribution of the short burst between the F and the P spectra is different. The K-S test gives a significance probability $8.40 \times 10^{-45}$ and $D$ value 0.29, which also shows that the distribution of the long burst between the F spectra and the P spectra is also different.

*3.2. The Correlation Properties between the Spectral Widths and the Model-independent Physics Quantities*

Both Paper I and our results above seem to show some differences between the long and short GRBs. Longer bursts seem to have relatively wider spectra. Paper II also pointed out that the photon fluence is correlated with the duration of a burst. These motivate us to suspect that the spectral width may relate to duration and other physics quantities. In this section, we first examine the correlated relationship between the spectral width and the model-independent physics quantities ($T_{90}$ duration, the 64/256/1024 ms peak flux and fluence in the 10–1000 keV energy range) for the entire burst set. We then check if there exist any differences between the long and short bursts for these correlated relationships. The correlations between the properties of the GRB spectra with spectral width examined in this study exhibit many correlations. The main correlated characteristic for the entire burst, long burst, and short burst are listed in Table 4.

Figure 2 demonstrates the spectral width $W$ versus duration for the P spectra and F spectra. When considering the entire burst set the correlated relationships (correlation coefficient $R = 0.29$, 0.40 and $p = 6.46 \times 10^{-23}$, ∼0.00 for the P spectra and the F spectra, respectively) are identified. This seems to indeed show that longer duration has a wider spectrum. However, the correlation properties of the entire, long, and short burst set are very different. Although they are correlated for the long burst set, the correlation is less significant than that of the entire burst set. While for the short burst set there is very weak correlation (see, Table 4), the short bursts extend the correlated trend for the long ones.

Burst fluence also correlates with $W$ for the entire burst set, both for the P and the F spectra, shown in Figure 2. This means that a wider burst has larger fluence. However, similar to the case of $W$ versus duration the correlations are also very different for the long and short bursts. The correlation of the long burst set is also much weaker than that of the entire burst set and no evident correlation is seen in short bursts for both spectra, but the short bursts also extend the correlated trend for long ones.

Figure 2 also demonstrates the relationships between $W$ and peak flux in 1024, 256, and 64 ms timescale for two classes of spectra. For both spectra, the spectral width positively correlates with the peak flux for all of the cases for the entire burst set (see, Table 4). These seem to mean that the brighter bursts (as measured by peak flux) are bursts with wider spectra. Different from the two cases above, the correlations of the long burst set are even stronger than those of entire burst set for all three timescales for both spectra, but there is no correlation for the short burst set. Moreover, the correlation and significance are not the same for the three different timescales and different spectra. The correlation of the F spectra is much stronger than that of the P spectra. For the entire burst set, the correlation and significant decrease with the decrease of the timescale and the peak flux with 1024 ms timescale show the strongest and statistically most significant correlation for both F spectra and P spectra.

From all the scatter plots above we have found that the dispersions around the correlation are great. The reason for this is that most of the widths are from the COMP model with smaller values and the number of wider spectra is much smaller than the others (see Figure 2 and Tables 1 and 2).

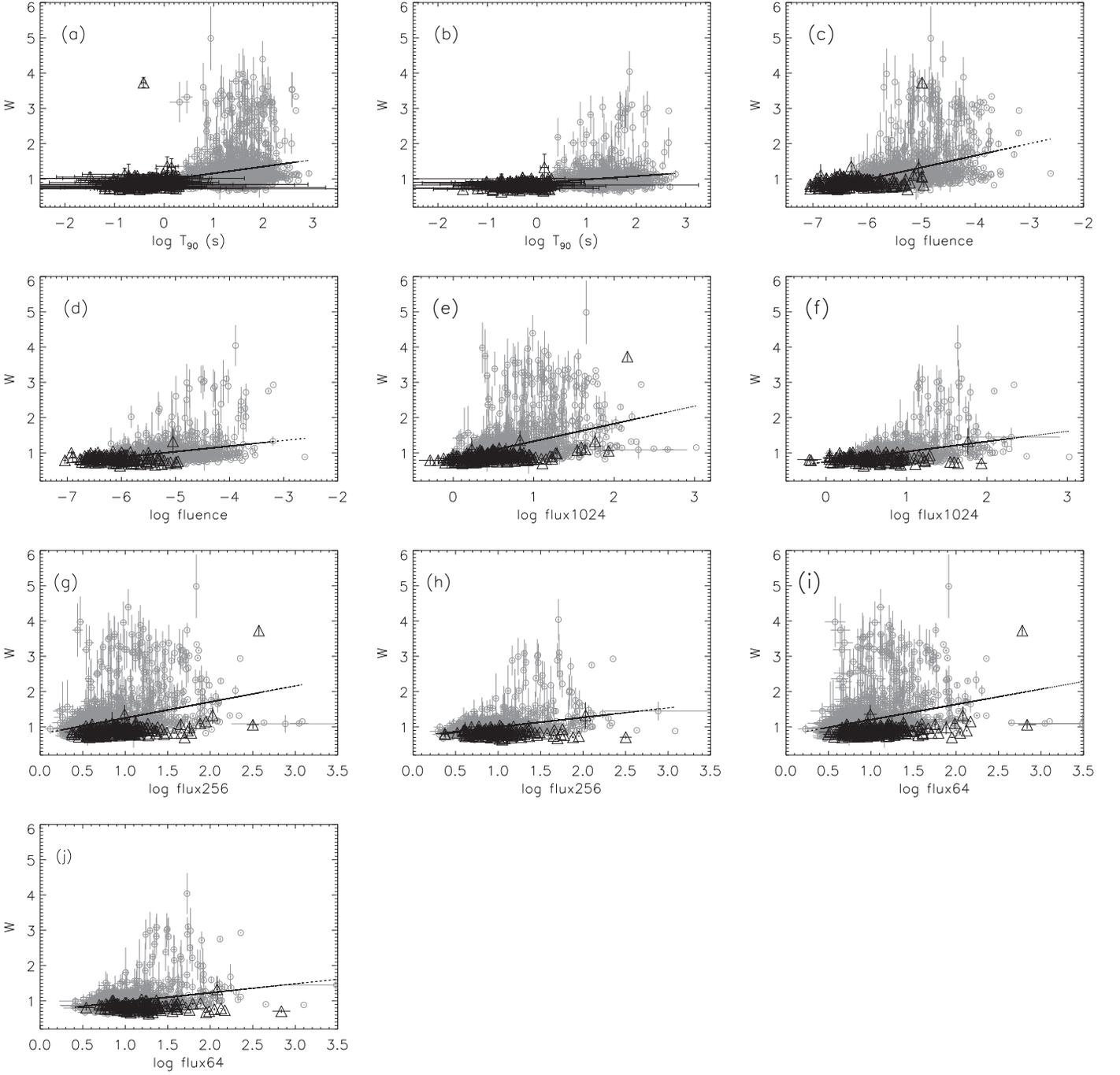

**Figure 2.** Spectral width $W$ vs. duration $T_{90}$ for the F spectra (a) and the P spectra (b), $W$ vs. fluence for the F spectra (c) and the P spectra (d), $W$ vs. flux in 1024 ms timescale for the F spectra (e) and the P spectra (f), $W$ vs. flux in 256 ms timescale for the F spectra (g) and the P spectra (h), $W$ vs. flux in 64 ms timescale for the F spectra (i) and the P spectra (j), where the triangles and the solid lines are the short bursts and the best-fitting lines for the entire burst set.

### 3.3. The Intrinsic Connection between the Spectral Width and Isotropic Radiated Energy or Isotropic Peak Luminosity

The previous section reveals some interesting statistically correlated relationships, indicating that the spectral width is related to duration, fluence, and peak flux. We wonder if these correlations are observed properties or intrinsic ones? Therefore, we investigate this issue with a sample containing redshift. The sample with known redshift consists of 75 and 86 bursts for P and F spectra, in which there are two and five short bursts, respectively. The isotropic energy $E_{\rm iso}$ and luminosity, $L_{\rm iso}$ (erg s$^{-1}$ or photons s$^{-1}$), that corresponds to a certain energy, observed from a source at a redshift, $z$, is given by

$$D_L = \frac{(1+z)c}{H_0} \int_0^z \frac{dz'}{\sqrt{(1+z')^2(1+\Omega_M z') - z'(2+z')\Omega_\Lambda}} \quad (3)$$

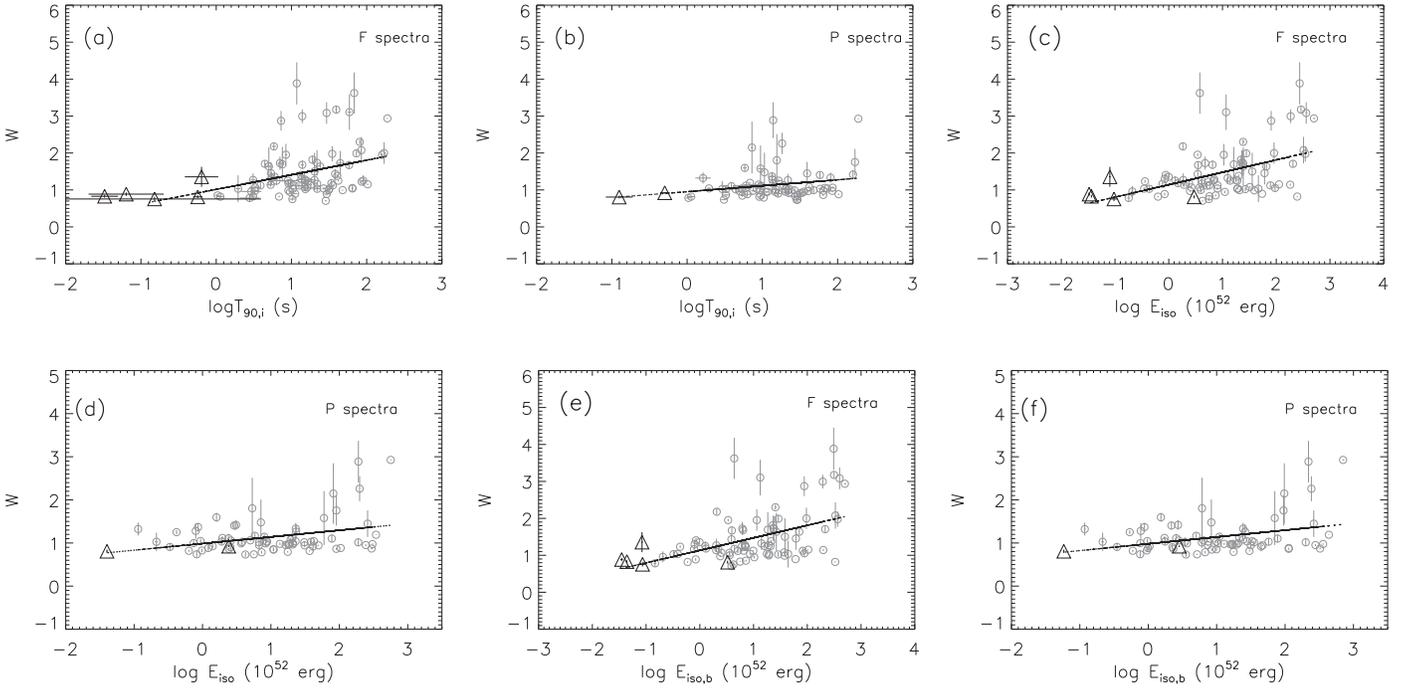

**Figure 3.** Spectral width $W$ vs. intrinsic duration $T_{90,i}$, isotropic energy $E_{iso}$ for F spectra and P spectra, where the subscript b denotes the BATSE energy channel, the triangles and the solid lines denote short bursts, and the best-fitting lines denote the entire burst set.

$$k = \frac{\int_{\frac{1}{1+z}}^{\frac{10^4}{1+z}} EN(E) dE}{\int_{E_{min}}^{E_{max}} EN(E) dE} \quad (4)$$

$$E_{iso} = \frac{4\pi S}{1+z} d_L^2 \times k \quad (5)$$

$$L_{iso} = \frac{4\pi F}{1+z} d_L^2 \times k, \quad (6)$$

where $d_L$ and $H_0$ are the luminosity distance and the Hubble constant, $\Omega_M$ is associated with the present day matter density, $\Omega_\Lambda$ with the dark energy density. $H_0$, $\Omega_M$, and $\Omega_\Lambda$ are taken as 70, 0.3, and 0.7, respectively. In this way, we can compute the isotropic energy and luminosity in the $E_{min}$–$E_{max}$ energy band. The intrinsic duration $T_{90,i}$ is $T_{90}/(1+z) \times K$. In this section, we consider the fluence and peak flux in the 10–1000 and 50–300 keV energy band to decrease the selection effect.

Figure 3 demonstrates the relations between the spectral width and intrinsic duration for P (a) and F spectra (b). It is found that the correlated properties are very different for different spectra. A correlation with the Spearman rank correlation coefficient $R = 0.41$ ($p = 0.01\%$) is identified for F spectra. Moreover, the short burst set significantly extends the correlated trend (also see Table 5). However, a Spearman rank correlation analysis shows that this correlation is not significant ($R = 0.15$, $p = 0.19$) for P spectra. These seem to show that the correlated property is only found for burst behavior.

Figure 3 also illustrates the spectral width versus burst isotropic energy for the 10–1000 and 50–300 keV energy bands. Similar to the case of $W$ versus burst intrinsic duration, a clear correlated relationship between $W$ and total radiated energy in both energy bands ($R = 0.41$, $p = 0.01\%$, and $R = 0.43$, $p = 0.004\%$ for 10–1000 keV and 50–300 keV, respectively) are also found for F spectra and are very weak, even having no correlation, for P spectra (also see Table 5).

Checking the relationship between the spectral width and the isotropic peak luminosity we also find that the $W$ is only correlated with the peak luminosity of F spectra for the 1024, 256, and 64 ms timescale (see, Figure 4 and Table 5). From Table 5, we also find that the correlation of 1024 ms timescale in 10–1000 energy band is the most significant.

In all the correlated cases above, the short bursts extend the correlation in a lower value area (see Table 5). The correlations are not evident when only considering the long bursts. These properties are very similar to the case of the observer frame.

### 4. The Dispersion Analysis and the Effect of the GBM Instrument on the Spectral Width

Comparing the Spearman correlation coefficients between $W$ and $E_{iso}$ as well as peak luminosity with published tables we found that they are much higher than the critical Spearman correlation coefficients at a 5% significance level (∼0.220 with dof = 77). It appears to be statistically weaker from the Spearman correlation analysis. In fact, there are several known correlation coefficients based on different statistical hypotheses. The Pearson correlation coefficient is the first formal correlation measure and used most often to describe the relationship between two variables. We also compute the Pearson correlation coefficient and the $p$-value ($r = 0.47, 0.38, 0.35, 0.32$, $p = 5.54 \times 10^{-6}, 3.03 \times 10^{-4}, 9.71 \times 10^{-4}, 2.40 \times 10^{-3}$ for $W$–$E_{iso}$, $W$–$L_{1024}$, $W$–$L_{256}$, and $W$–$L_{64}$, respectively) and also reveals a weaker correlation.

Goodwin & Leech (2006) described and illustrated six factors that affect the size of a Pearson correlation, which include differences in the shapes of the two distributions, the presence of "outliers," the amount of variability in either variable, lack of linearity, characteristics of the sample, and measurement error. We mainly consider the first three factors in

Table 5
Spearman Rank Correlation Analysis Results of the Spectral Widths and the Intrinsic Model-independent Physics Quantities

| Parameter Pairs | P Spectra Entire Bursts | Long GRBs | F Spectra Entire Bursts | Long GRBs |
|---|---|---|---|---|
| $W - T_{90,i}$ | 0.15 ($1.93 \times 10^{-1}$) | 0.10 ($3.86 \times 10^{-1}$) | 0.41 ($1.01 \times 10^{-4}$) | 0.34 ($1.98 \times 10^{-3}$) |
| $W - E_{\rm iso}$ | 0.16 ($1.78 \times 10^{-1}$) | 0.12 ($3.20 \times 10^{-1}$) | 0.41 ($1.10 \times 10^{-4}$) | 0.35 ($1.19 \times 10^{-3}$) |
| $W - E_{\rm iso,b}$ | 0.17 ($1.55 \times 10^{-1}$) | 0.13 ($2.72 \times 10^{-1}$) | 0.43 ($4.38 \times 10^{-5}$) | 0.38 ($5.14 \times 10^{-4}$) |
| $W - L_{1024}$ | 0.09 ($4.32 \times 10^{-1}$) | 0.06 ($6.35 \times 10^{-1}$) | 0.32 ($2.54 \times 10^{-3}$) | 0.27 ($1.44 \times 10^{-2}$) |
| $W - L_{1024,b}$ | 0.10 ($4.03 \times 10^{-1}$) | 0.07 ($5.37 \times 10^{-1}$) | 0.32 ($1.71 \times 10^{-3}$) | 0.28 ($1.23 \times 10^{-2}$) |
| $W - L_{256}$ | 0.08 ($4.88 \times 10^{-1}$) | 0.06 ($6.64 \times 10^{-1}$) | 0.30 ($6.19 \times 10^{-3}$) | 0.25 ($2.28 \times 10^{-3}$) |
| $W - L_{256,b}$ | 0.09 ($4.46 \times 10^{-1}$) | 0.07 ($5.56 \times 10^{-1}$) | 0.29 ($7.08 \times 10^{-3}$) | 0.27 ($1.65 \times 10^{-2}$) |
| $W - L_{64}$ | 0.08 ($5.16 \times 10^{-1}$) | 0.05 ($6.59 \times 10^{-1}$) | 0.27 ($1.26 \times 10^{-2}$) | 0.24 ($3.29 \times 10^{-2}$) |
| $W - L_{64,b}$ | 0.08 ($5.15 \times 10^{-1}$) | 0.06 ($6.24 \times 10^{-1}$) | 0.26 ($1.47 \times 10^{-3}$) | 0.25 ($2.72 \times 10^{-2}$) |

**Note.** $L_{1024}$, $L_{256}$, and $L_{64}$ are the peak luminosity in 1024, 256, and 64 ms timescales in the 10–1000 keV energy band, and "b" stands for the BATSE (50–300 keV) energy band.

this paper. We first investigate the distribution shapes of two variables. If the two distributions have more similar shapes, the correlation has a higher maximum value. The distributions of $W$, $E_{\rm iso}$, and $L_{1024}$ presented in Figure 5 show that the shape of $W$ is slightly different from those of $E_{\rm iso}$ and $L_{1024}$. The skew distribution of $W$ may be caused by too many smaller $W$ values (COMP model) or too few wider spectral bursts (BAND and SBPL model).

We then check if there are regression "outliers" between $W$ and $E_{\rm iso}$, $L_{1024}$, $L_{256}$, and $L_{64}$. There are several methods to test regression "outliers." We employ the linear model with both errors in $X$ and $Y$ data to estimate the best-fitting lines with the Hyper-fit package (Robotham & Obreschkow 2015). The associated residuals and intrinsic scatters are then obtained. We select the outer ~30% of the data as outliers due to the great scatters of the relationships. That is to say, we only consider that data within one intrinsic scatter and recheck their correlation coefficients. The expected correlation properties are listed in Table 6. It is found that all the correlation coefficients are above 0.60. As we can see from Figure 6, the outlier bursts are in the $W$-direction, those brighter bursts with wide spectra or those less bright but with very small $W$.

As Goodwin & Leech (2006) pointed out, the value of correlation coefficient will be lower if there is less variability among the observations than if there is more variability. Note that the variability here is the amount between the smallest and the largest item in the data set. So we suspect that the skew distribution of $W$ is caused by the definition of $W$. We try to adopt the absolute width because the values of relative spectral width $W$ in the rest frame cancel out the correction of redshift. The variability of the absolute width $W_{ab} * (1 + z)$ (4.22) in the rest frame indeed increases comparing with that of the relative width (3.18). Using the absolute width, we find in Figure 7 and Table 7 that the correlations are much tighter. In addition, it is found that the correlations also exist for the P spectra. Therefore, it is shown that there are indeed correlated relationships between the spectral width and the isotropic energy, as well as the peak luminosity. However, the scatters of the correlated relationship are still great, especially for the F spectra.

Therefore, we suspect that the inaccurate measure of model-dependent spectral width causes these great scatters. The spectral widths in our sample are obtained with the BEST models, but the BEST spectra are only the best estimates of the observed GRB spectra by using model comparison. The fitted photon models are mainly dependent on the photon number above the background received from the detectors. In fact, the detector effective area and the photon flux fall very rapidly with increasing photon energy. Therefore, the detected high-energy photons are very few for those dim bursts. Even if for the brighter bursts the high-energy photon number detected may be much lower than the actual ones. Consequently, it is very difficult to find the true photon models because there must be some fit at which photon model spectra cannot represent true burst spectra.

In fact, Band et al. (1993) have realized this problem and they found that the simulated BAND spectra with low signal-to-noise ratio (S/N) could be adequately fitted with the three-parameter COMP model. Kaneko et al. (2006) also investigate this issue with BATSE data and the simulated results indicated that S/N ~ 80 is needed for the BAND fits to be better than the COMP fits, at the ~99.9% confidence level. For spectra with S/N ~ 40, the confidence level of improvements in BAND over COMP were <70%. But nobody has done similar work for GBM. So we also simulate a BAND spectrum in a GBM detector at various S/N, fit it with a COMP function, and determine at what confidence level a COMP would be favored instead of the BAND.

The minimum S/N of our sample spectra is set to 3.5 and is much smaller than BATSE sample provided by Kaneko et al. (2006). In addition, GBM has slightly worse sensitivity than BATSE. So we deduce that the confidence level of BAND over COMP would be much less than 70%. We investigate the effect of the BEST BAND model on spectral parameters and on our analysis results. We select four bursts in our sample with the BEST BAND model and defined a set of simulations using Band as the input model (i.e., null hypothesis). We create a set of 10,000 simulated GRB spectra with an actual BAND spectral parameter and each spectrum covering the same duration as the real source time interval at various S/N. For the S/N variation, we use the actual fitted amplitude values of 3.5, 10, 30, 50, 70, 90, and 120. The background counts of the synthesized spectra are estimated from the real data. The input source counts are then folded through the DRM. Finally, Poisson noise was added to the sum of the source and background counts. The synthetic spectra were then fitted with BAND and COMP models, adding Poisson fluctuations to each energy channel of the background spectrum during the fit process. We use the newest version of RMFIT to perform all simulations. Applying the BEST model criteria $\Delta C - {\rm Stat}_{\rm crit} = 11.83$ for COMP versus BAND we can determine at

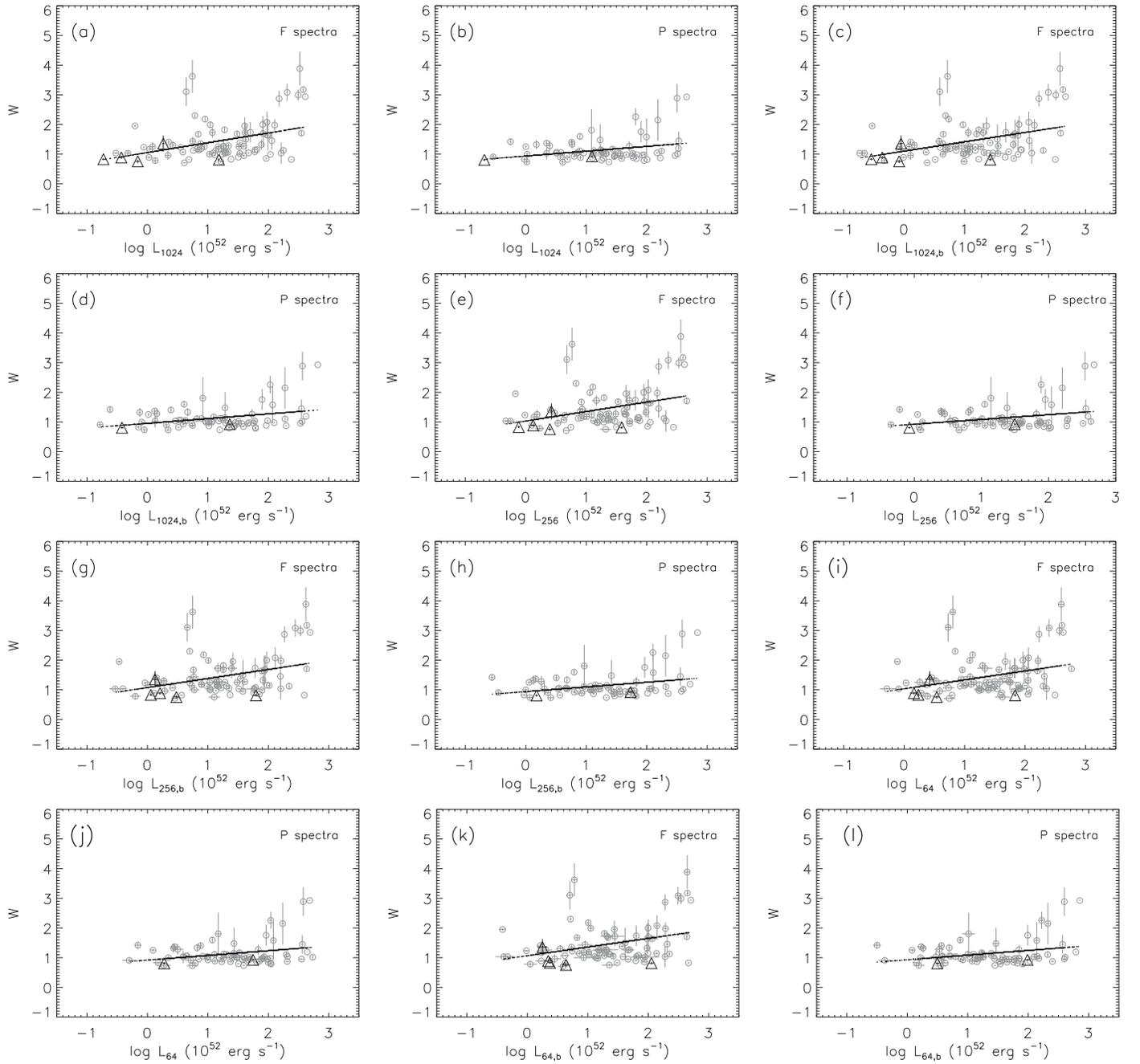

**Figure 4.** Spectral width $W$ vs. isotropic peak luminosity in 1024, 256, 64 ms timescale for F spectra and P spectra, where the subscript b denotes the BATSE energy channel, the triangles and the solid lines are the short bursts and the best-fitting lines for entire burst set.

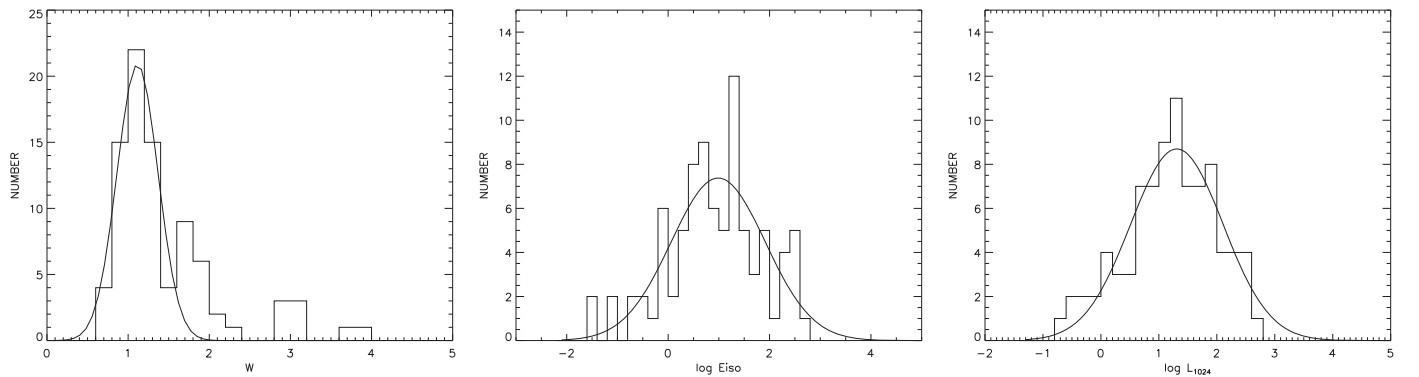

**Figure 5.** Distributions of $W$, $E_{\mathrm{iso}}$, and $L_{1024}$ for F spectra.

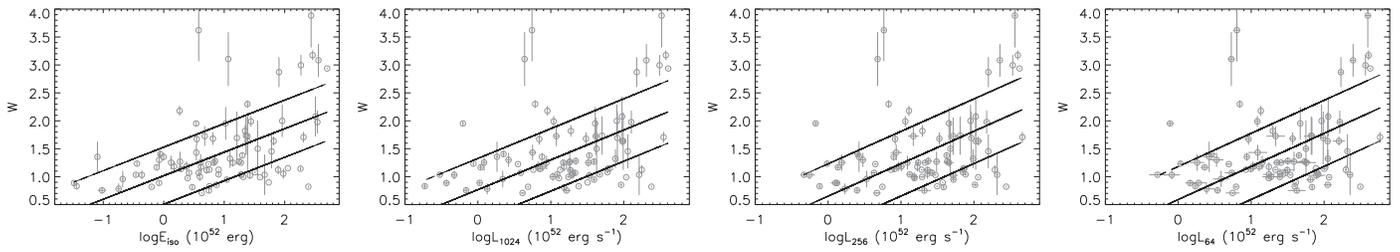

**Figure 6.** Spectral width vs. isotropic energy, peak luminosity in 1024 ms, 256 ms, and 64 ms timescale for F spectra, where the solid lines and the dashed lines are the best-fit lines and the $\pm 1\sigma$ dispersion region of the correlations, respectively.

**Table 6**
Correlation Analysis Results of the Spectral Widths and the Isotropic Energy as well as the Peak Luminosity after Removing the Outliers

| Parameter Pairs | Pearson | Spearman Rank |
|---|---|---|
| $W - E_{iso}$ | 0.73 ($1.78 \times 10^{-11}$) | 0.70 ($8.60 \times 10^{-10}$) |
| $W - L_{1024}$ | 0.66 ($4.32 \times 10^{-9}$) | 0.65 ($2.45 \times 10^{-8}$) |
| $W - L_{256}$ | 0.65 ($4.03 \times 10^{-8}$) | 0.65 ($5.70 \times 10^{-8}$) |
| $W - L_{64}$ | 0.63 ($1.55 \times 10^{-7}$) | 0.61 ($4.10 \times 10^{-7}$) |

**Note.** The $L_{1024}$, $L_{256}$, and $L_{64}$ are the peak luminosity in 1024, 256, and 64 ms timescale in 10–1000 keV energy band.

**Table 7**
Spearman Rank Correlation Analysis Results of the Absolute Spectral Width and the Intrinsic Energetics as well as Peak Luminosity

| Parameter Pairs | F Spectra | P Spectra |
|---|---|---|
| $E_{iso} - W_{ab,i}$ | 0.68 ($6.05 \times 10^{-13}$) | 0.65 ($2.80 \times 10^{-10}$) |
| $L_{1024} - W_{ab,i}$ | 0.56 ($2.58 \times 10^{-8}$) | 0.53 ($8.88 \times 10^{-8}$) |
| $L_{256} - W_{ab,i}$ | 0.57 ($1.30 \times 10^{-8}$) | 0.55 ($2.88 \times 10^{-7}$) |
| $L_{64} - W_{ab,i}$ | 0.57 ($1.55 \times 10^{-8}$) | 0.57 ($1.71 \times 10^{-7}$) |

what confidence level a COMP would be favored instead of the BAND.

Our F spectral sample contains 168 GRBs with the widths fitted by the BEST BAND models. Similar to Paper I we use four typical bursts located in two different energy fluence classes, which include one of the brightest bursts and one of the dimmest bursts. The simulation results are presented in Table 8. From Table 8 the number of favor for the COMP model instead of the Band model is very different for different fluence classes and the number seems to increase with the decrease of the fluence. The difference of the $C - STAT$ between the COMP and BAND models increases with the increase of the brightness. We can also find that none of the simulated spectra are best fitted by the COMP model regardless of S/N when we simulate the brightest burst among our sample. The favored number of the COMP model of the dimmest burst is very large. We use the median number for the favoring COMP model selection of all simulated bursts. Moreover, it is shown in Table 8 that the number of the favored COMP model does not vary with the S/N. Based on the median number, our simulation result shows an ~46% confidence level of improvements in COMP over BAND regardless of S/N. This result seems to show that the spectral parameters are affected by the spectral brightness and the dimmer the bursts, the less accurate the spectral parameters. Therefore, there must be some outliers in the spectral parameters and it must affect our observed spectral width. The magnitude of the effect is related to the burst fluence. But we cannot identify which bursts are outliers in our sample from simulations.

## 5. Discussion and Conclusions

We have reanalyzed the width of GRB spectra with BEST model parameters from *Fermi*/GBM for time-integrated and peak flux spectra, respectively. Different from Paper I, the BEST spectral data rather than Band model alone are used. The majority of BEST spectral data are fitted by the Compton model (888/967, 1228/1431 for the P and F spectra, respectively). Only a small portion of bursts is fitted by Band (57/967, 135/1431 for the P and F spectra, respectively) and SBPL (22/967, 68/1431 for the P and F spectra, respectively) model. The best estimate of the observed properties of GRBs based on the data received are obtained for the BEST parameter sample. The data from different GRBs tend to support different empirical models and it will be possible to determine the physics of the emission process by investigating the tendencies of the data to support a particular model over others (see Paper II).

We first compute the BEST widths for peak flux and time-integrated spectra to compare the results provided by Paper I. When comparing the P spectral width our median values $0.96 \pm 0.09$ and $0.82 \pm 0.10$ for long and short bursts are much smaller than the corresponding median values of Paper I because most of the spectral widths are computed by the Compton model. The median width of the long burst is $0.96 \pm 0.09$ and much greater than that of the short burst median width $0.82 \pm 0.10$ for the P spectra. While for the F spectra the median values of long and short bursts are $1.09 \pm 0.05$ and $0.88 \pm 0.09$, respectively. These medians seem to show that the short burst spectra are different from that of long bursts. Moreover, the K-S test of the long and short burst set shows that they come from different distribution for both spectra. However, the distributions of long and short bursts are perfectly compatible when taking into account the variances of the distributions.

Comparing the observed width with the width of thermal emission computed by Paper I we can also confirm that the observed spectra cannot be interpreted by Planck function alone based on the fact that no value is less than 0.59 within our estimated uncertainty for two types of spectra. Similar to Paper I both the median widths ($1.05 \pm 0.04$ for F spectra and $0.94 \pm 0.08$ for P spectra) are close to the width of monoenergetic synchrotron (0.93) but this is not a physically model. Synchrotron emission widths from all electron distributions are around 1.5 and much wider than the observed median value, which seems to show synchrotron emission can only interpret those wider observed long burst spectra. While for short bursts few observed spectra are wider than 1.4 within

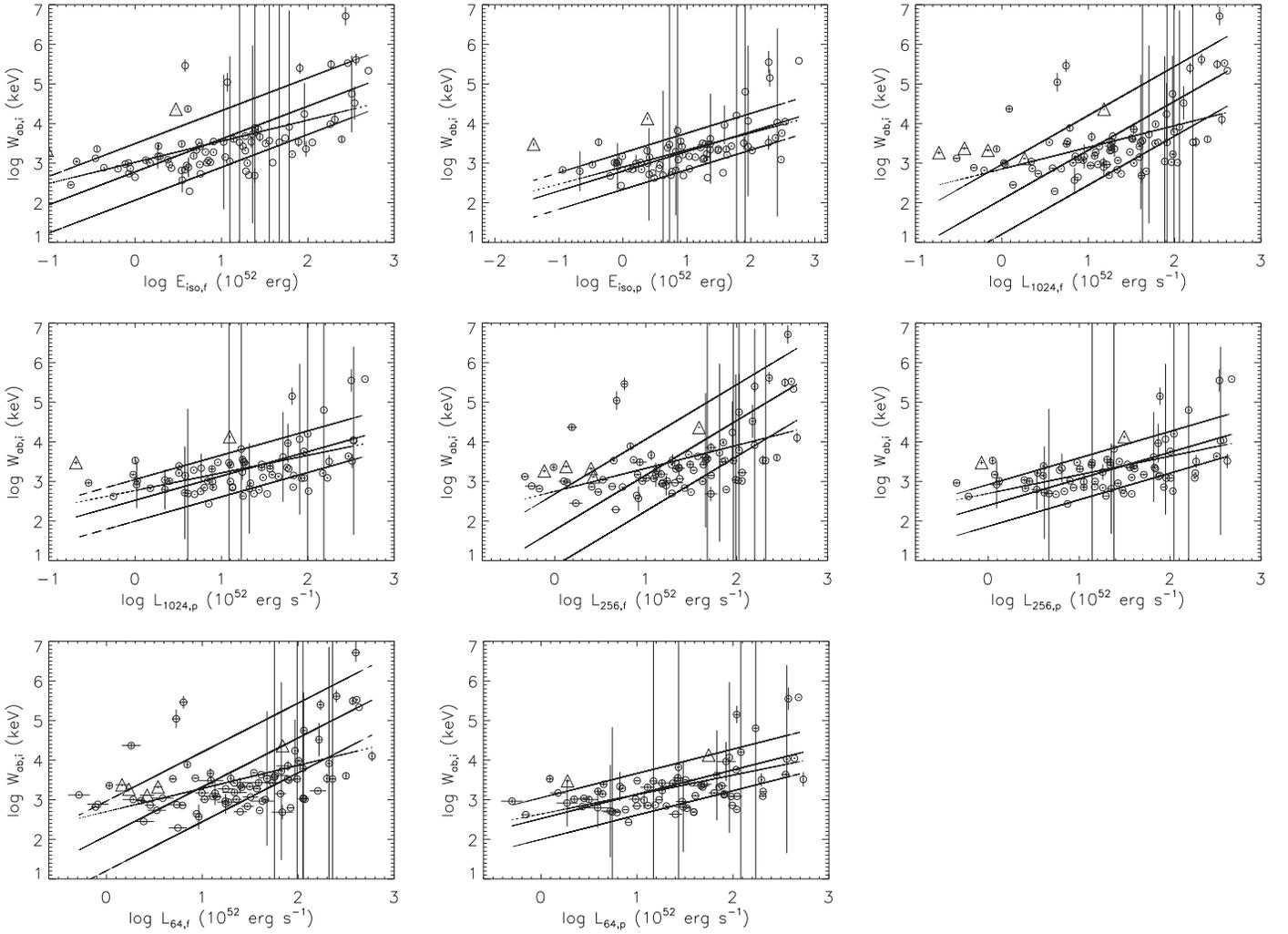

**Figure 7.** Intrinsic absolute spectral width $W_{ab,i}$ vs. isotropic energy, peak luminosity in 1024 ms, 256 ms, 64 ms timescales for F spectra and P spectra, where the subscripts f and p denote F spectra and P spectra, respectively. The triangles represent the short bursts. The dotted lines and solid lines correspond to the best-fitting lines without and with errors for the long and short bursts. The dashed lines are the $\pm 1\sigma$ dispersion region of the correlations.

**Table 8**
The Fit Results to Simulated BAND Spectra

| GRB name | $N(3.5)$ | $N(10)$ | $N(30)$ | $N(50)$ | $N(70)$ | $N(90)$ | $N(120)$ | Fluence | $\Delta C - \mathrm{STAT}^a$ |
|---|---|---|---|---|---|---|---|---|---|
| GRB171010792 | 0 | 0 | 0 | 0 | 0 | 0 | 0 | $6.33 \times 10^{-4}$ | 859 |
| GRB120711115 | 3336 | 3311 | 3273 | 3447 | 3408 | 3412 | 3422 | $1.94 \times 10^{-4}$ | 62 |
| GRB170705115 | 5864 | 5959 | 5659 | 5755 | 5632 | 5566 | 5474 | $1.34 \times 10^{-5}$ | 20 |
| GRB081222204 | 8073 | 7804 | 7944 | 7807 | 7971 | 7749 | 7979 | $1.19 \times 10^{-5}$ | 18 |
| $\langle N \rangle^b$ | 4600 | 4635 | 4466 | 4601 | 4520 | 4489 | 4448 | | |

**Notes.** The $N$ is the number favoring the COMP model and the corresponding values of S/N are shown in parentheses.
[a] The difference in $C - \mathrm{Stat}$ between the COMP and BAND.
[b] The median of the number favoring the COMP model.

our estimated uncertainties. These reveal that most of the observed spectra are much narrower than the emission from an electron distribution and are also significantly wider than a blackbody spectrum. These appear to suggest that blackbody and synchrotron emission alone cannot explain most of the observed spectrum. However, our observations show that there are significant thermal spectral components in *Fermi* GRBs (e.g., Yu et al. 2016) and the synchrotron emission models are thought to be a popular model to interpret GRB prompt emission spectra. Therefore, the observed spectra must involve other mechanisms. Recently, Bharali et al. (2017) suggested that if the system undergoes a rapid temperature evolution the observed spectral shape can be broadened. Particularly, if invoking thermal radiation, a way to broaden the spectrum must be found. Paper I employed a deep discussion on this issue; thus, one can refer to Paper I for detailed information.

Using these BEST model spectral widths we also investigate the relationships between the spectral width and the model-

**Table 9**
Spearman Rank Correlation Analysis Results of the Correlated Index and the Intrinsic Physics Quantities

| Parameter Pairs | P Spectra | F Spectra |
|---|---|---|
| index–$W$ | $-0.90$ ($1.97 \times 10^{-28}$) | $-0.71$ ($1.27 \times 10^{-14}$) |
| index–$t_{90,i}$ | $-0.10$ ($3.81 \times 10^{-1}$) | $-0.24$ ($2.67 \times 10^{-2}$) |
| index–$E_{\rm iso}$ | $-0.25$ ($3.43 \times 10^{-2}$) | $-0.56$ ($1.64 \times 10^{-8}$) |
| index–$E_{\rm iso,b}$ | $-0.25$ ($2.87 \times 10^{-2}$) | $-0.59$ ($2.87 \times 10^{-9}$) |
| index–$L_{1024}$ | $-0.16$ ($1.69 \times 10^{-1}$) | $-0.49$ ($1.35 \times 10^{-6}$) |
| index–$L_{1024,b}$ | $-0.18$ ($1.24 \times 10^{-1}$) | $-0.50$ ($9.91 \times 10^{-7}$) |
| index–$L_{256}$ | $-0.15$ ($2.12 \times 10^{-1}$) | $-0.46$ ($7.03 \times 10^{-6}$) |
| index–$L_{256,b}$ | $-0.17$ ($4.40 \times 10^{-1}$) | $-0.48$ ($2.61 \times 10^{-6}$) |
| index–$L_{64}$ | $-0.14$ ($2.36 \times 10^{-1}$) | $-0.44$ ($2.69 \times 10^{-5}$) |
| index–$L_{64,b}$ | $-0.15$ ($1.88 \times 10^{-1}$) | $-0.46$ ($1.08 \times 10^{-5}$) |

**Note.** The $L_{1024}$, $L_{256}$, and $L_{64}$ are the peak flux in 1024, 256, and 64 ms timescale in 10–1000 keV energy band and "b" stands for the BATSE (50–300 keV) energy band.

independent physics quantities. The correlations between the spectral width and the burst duration, fluence, and peak flux in 1024, 256, and 64 ms timescale appear to show that the spectral width is an important quantity. However, the correlations of the entire burst, long, and short burst sets are different (see, Table 4). The correlations of the long burst set are weaker than that of the entire burst set and there have been no evident correlations for the short burst set. These seem to show that there are two different mechanisms capable of producing GRB correlated properties.

Whether these correlations are intrinsic or not is still uncertain. We select 75 P bursts and 86 F bursts with known redshift to verify it. It is found that $W$ is correlated with intrinsic duration for F spectra but there is no correlation for P spectra. It is very interesting that $W$ is correlated with $E_{\rm iso}$ and $L_{\rm iso}$ in both energy ranges 10–1000 and 50–300 keV. These suggest that the spectral width is related to energy and luminosity. Moreover, we find that short bursts still extend the correlation for long bursts. However, the correlations are not identified for P spectra. The possible reason for this is that there are too few counts with such short times (1024 ms for long burst and 64 ms for short burst) used to determine the peak flux spectrum.

We also investigate the possible reasons of the different distributions of the long and short bursts and the correlations between the spectral width and the other parameters. Because the sample consists of BAND, COMP, and SBPL, the spectral widths also consist of the three models. For the Band model, the widths are strongly correlated with the high-energy index $\beta$ ($R=0.93$, $p=1.09 \times 10^{-7}$) and less correlated with $\alpha$ ($R=-0.60$, $p=1.30 \times 10^{-2}$) as well as $\alpha - \beta$ ($r=-0.88$, $p=6.10 \times 10^{-6}$). While for the SBPL model, the widths are also strongly correlated with the high-energy index $\lambda_2$ ($R=0.97$, $p=5.14 \times 10^{-9}$) and less correlated with $\lambda_1$ ($R=-0.42$, $p=1.16 \times 10^{-1}$) as well as $\lambda_1 - \lambda_2$ ($R=-0.89$, $p=7.49 \times 10^{-6}$). It is very evident that the widths are strongly correlated with $\alpha$ ($R \sim -1$, $p \sim 0.00$) for the COMP model. Here we call the high-energy index $\beta$, $\lambda_2$ and the low-energy index $\alpha$ the correlated index. Therefore, we first suspect that these intrinsic correlations may result from the correlations between these correlated indexes and those intrinsic physics quantities. Therefore, we compute the statistical significance of the correlations between these intrinsic physics quantities and the correlated index, which are also listed in Table 9. From Table 9 we can find that the correlations between these intrinsic physics quantities and the correlated index are much weaker than those of $W$ versus the correlated index. Therefore, we think the correlations between $W$ and the intrinsic physics quantities may not be caused by the correlations between these intrinsic physics quantities and correlated index and there must be another factor that causes the correlations. The strong dependence of widths on the low-energy index $\alpha$ can interpret why the short bursts with relatively narrower spectra are due to the fact that short GRBs are harder than long ones and the majority of BEST spectral data are fitted by the COMP model.

Our sample only includes the GBM data, covering an energy range from $\sim 8$ keV to $\sim 38$ MeV. When taking the LAT data into account to determine if the correlated relationships would change, Paper II has pointed out that the median value from the GBM-only fits decreased from $\beta \sim -2.2$ to $-2.5$ for the GBM and LAT joint spectral fits. That is, the expected high-energy index would be steeper for the GBM and LAT joint spectral fits. Therefore, the spectral widths would be narrower due to the fact that the spectral width is positively correlated with the high-energy index. Correspondingly the expected flux in the GBM and LAT joint spectral fits is lower than that predicted by the GBM-only spectral shape (see Massaro et al. 2010). In this way the correlated relationships between the spectral widths and the energy as well as peak luminosity still exist.

In fact, the GBM is only an observational instrument and has less sensitivity than BATSE. The observed data must be affected by the energy range, effect area, sensitivity, etc. The GBM detector effective area and the flux fall rapidly with increasing energy, which leads to the photon models being inaccurate for the weaker GRBs. Therefore, it is very difficult to determine which model is statistically more preferable. This must lead to the inaccurate measure of spectral parameters. We have also investigated the effect of the GBM instrument on our measured spectral widths. Our simulation result shows that an $\sim 54\%$ confidence level of improvement in BAND over COMP regardless of S/N. Moreover, the confidence level does not change with the S/N. The lower confidence level causes a higher inaccurate measure of the spectral width and may lead to less strong correlations between the spectral widths and the intrinsic physics quantities.

Our sample bursts are only fitted with a model or component. However, Guiriec et al. (2010, 2011, 2013, 2015a, 2015b, 2016a, 2016b) have pointed out that some prompt emission GRB spectra are composed of a superposition of several components, such as a thermal spectral component (BB) and PL. If there is an unresolved thermal component or other component in our sample bursts that modifies the non-thermal-only fit parameters, then we might expect a systematic bias yielding values of $\alpha$ and $\beta$. They found that these components affecting the parameters of the Band function or COMP are similar for different bursts: both $\alpha$ and $\beta$ are shifted toward lower values. In other words both $\alpha$ and $\beta$ are greater than their real values. Taking the GRB 100724B, for example, $\beta$ changes from $\sim -2$ (Band-only) to $-2.11$ (Band+BB) and $-2.13$ (Band+Comp) (see Table 1 in Guiriec et al. 2011). This leads to the decreases in spectral width. Moreover, as we can see, the flux also decreases when an addition component is fitted (see, e.g., Figure 15 in Guiriec et al. 2015a and Figure 11 in Guiriec et al. 2013). Therefore, the positive correlation between the spectral widths and luminosity and energy are also identified.

Our results show another empirical connection between the measurable properties of the prompt gamma-ray emission and the luminosity or energy of GRBs is identified. Several similar empirical luminosity or energy correlations have been statistically found from observations in recent years. The first one is the well-known correlation between the peak energy of the $\nu F_\nu$ spectrum and the isotropic-equivalent energy ($E_p - E_{\rm iso}$, the Amati relation; Amati et al. 2002). The other correlations are the peak energy and collimation-corrected energy ($E_p - E_\gamma$; Ghirlanda et al. 2004) as well as the peak energy and isotropic peak luminosity ($E_p - L$; Schaefer 2003; Yonetoku et al. 2004) correlation. Moreover, all of these empirical correlations are based on the peak energy of time-integrated $\nu f_\nu$ spectra. While our empirical correlations are based on the spectral widths of time-integrated $\nu f_\nu$ spectra. Both the peak energy and the spectral widths are related to the spectral shape of GRB. Our study shows that the correlations between the spectral widths and energy or luminosity are identified for both spectra. It makes us suspect that there are some connections between the spectral shapes and energy as well as luminosity. Moreover, we also suspect that the other shape parameters (such as $E_2$, $E_1$) are correlated with luminosity as well as energy, and wonder which parameters are better indicators of energy and luminosity. We shall investigate these issues in detail in future work.

In conclusion our analysis results appear to show that the spectral widths are correlated with energy and peak luminosity in GRBs with known redshifts. If the relationships are confirmed with an increase in data from various instruments, we can estimate luminosities and energies by calibrating the spectral widths. In this way, a Hubble plot for GRBs can be obtained.

We would like to thank the anonymous referee for constructive suggestions to improve the manuscript. This research has made use of data, software, and/or web tools obtained from the High Energy Astrophysics Science Archive Research Center (HEASARC), a service of the Astrophysics Science Division at NASA/GSFC and of the Smithsonian Astrophysical Observatory's High Energy Astrophysics Division. This work was supported by the National Natural Science Foundation of China (grant Nos. 11763009, 11263006 and U1831135), the Yunnan Natural Science Foundation (2014FB188), the Natural Science Fund of the Education Department of Guizhou Province (KY2015455), and the Natural Science Fund of the Liupanshui Normal College (LPSSY201401).

### ORCID iDs

Z. Y. Peng  https://orcid.org/0000-0003-3846-0988